\documentclass[10pt,letterpaper]{article}
\usepackage[top=0.85in,left=2.75in,right=1in,bindingoffset=0in,footskip=0.75in,marginparwidth=2in]{geometry}
\usepackage[round,authoryear]{natbib}
\usepackage[utf8]{inputenc}
\usepackage{amssymb}
\usepackage{amsmath}
\usepackage{amsthm}

\usepackage{nameref,hyperref}

\usepackage[right]{lineno}

\usepackage{microtype}
\DisableLigatures[f]{encoding = *, family = * }

\raggedright
\usepackage{changepage}

\usepackage[aboveskip=1pt,labelfont=bf,labelsep=period,singlelinecheck=off]{caption}

\makeatletter
\renewcommand{\@biblabel}[1]{\quad#1.}
\makeatother

\usepackage{lastpage,fancyhdr,graphicx}
\usepackage{epstopdf}
\fancyheadoffset[L]{2.25in}
\fancyfootoffset[L]{2.25in}

\usepackage{color}

\definecolor{Gray}{gray}{.25}

\usepackage{graphicx}

\usepackage{sidecap}

\usepackage{wrapfig}
\usepackage[pscoord]{eso-pic}
\usepackage[fulladjust]{marginnote}
\reversemarginpar

\begin{document}
\vspace*{0.35in}

\begin{flushleft}
{\Large
\textbf\newline{Mutual gravitational energy of homogeneous prolate spheroids. Collinear case}
}
\newline
\\
Kondratyev B.P.\textsuperscript{1,2*}, 
Kornoukhov V.S.\textsuperscript{1}
Kireeva E.N.\textsuperscript{1}, 
\\
\bigskip
\bf{1} Sternberg Astronomical Institute, M.V. Lomonosov Moscow State University, 13 Universitetskij prospect, 119992, Russia
\\
\bf{2} Central Astronomical Observatory at Pulkovo, Russia 
\\
\bigskip
* work@boris-kondratyev.ru

\end{flushleft}\justify

\section*{ABSTRACT}
The problem of mutual gravitational energy $W_{mut}$ for a system of two homogeneous prolate spheroids, whose symmetry axes are on the same line, is set and solved. The method of equigravitating elements is applied, where the external potentials of three-dimensional spheroids are represented by the potentials of one-dimensional inhomogeneous focal rods. The solution of the problem is reduced to the integration of the potential of one rod over the segment of the second rod. As a result, the expression $W_{mut}$ for two prolate spheroids can be obtained in a finite analytic form through elementary functions. The force of attraction between the spheroids is found. The function $W_{mut}$ is also represented by a power series in eccentricity of the spheroids. Possible applications of the obtained results are discussed.\bigskip


\section{INTRODUCTION}\label{lab:intro}

To solve many physical and astronomical tasks it is needed to know gravitational (potential) energy $W$ of various bodies of different shapes. In its scalar form potential energy is necessary to calculate such gravitating system parameters as velocity dispersion and pressure, components of forces and torques. Potential-energy tensor is included in virial equations of the second order and is used in studying equilibrium and stability of celestial bodies's figures of equilibrium \citep{Chandra1973}.

Let us consider two masses $M_1$ and $M_2$, distributed in volumes $V_1$ and $V_2$, with densities $\rho_{1}$ and $\rho_{2}$ respectively. Each of these masses is being a source of a gravitational field with a potential
\begin{equation}\label{1}
 \varphi_{i} \left( x \right) = G \int \limits_{V_{i}} \frac{\rho_{i} \left( x \right) \ }{\vert x-x' \vert} dV, \quad i=1,2,
\end{equation}
and, being in the gravitational field of jth partner, has potential energy
\begin{equation}\label{2}
W_{i,j} = - \int \limits_{V_{i}} \rho_{i} \left( x \right) \varphi_{j} \left( x \right) dV.
\end{equation}

Gravitational energy is not regarded as an additive quantity. Thus, if some body (or a system of bodies) consists from, for example, two parts, its full gravitational energy may be written as
\begin{equation}\label{3}
\begin{array}{lcl}
 W = - \displaystyle \frac{1}{2} \left\lbrace \int \limits_{V_{1}} \rho_{1} \left( x \right) \varphi_{1} \left( x \right) dV +\int \limits_{V_{2}} \rho_{2} \left( x \right) \varphi_{2} \left( x \right) dV + \right. \\ \\
\left. + \int \limits_{V_{1}} \rho_{1} \left( x \right) \varphi_{2} \left( x \right) dV +\int \limits_{V_{2}} \rho_{2} \left( x \right) \varphi_{1} \left( x \right) dV \right\rbrace .
\end{array}
\end{equation}
Two first integrals in (\ref{3}) are equal to gravitational energies $W_1$ and $W_2$ for each of the separate subsystems, and the sum of two last integrals gives \textit{mutual gravitational energy} $W_{mut}$ for these two bodies (or parts of the body). Equality of these two integrals is a well-known property \citep{Kond1989}
\begin{equation}\label{4}
 W_{mut} = -\int \limits_{V_{1}} \rho_{1} \left( x \right) \varphi_{2} \left( x \right) dV = - \int \limits_{V_{2}} \rho_{2} \left( x \right) \varphi_{1} \left( x \right) dV.
\end{equation}
Thus, full gravitational energy of the entire body consists of three terms
\begin{equation}\label{5}
 W=W_{1}+W_{2}+W_{12}.
\end{equation}

Gravitational energy depends on the shape and internal structure of the body, so calculating $W$ is a complicated mathematical task, the solution of which rarely can be presented in a finite analitycal form. Usually in literature one can find expressions for the potential energy of spheres or ellipsoids only, look, for example, in \citep{Subb1936},~\citep{Chandra1973}. Recently a special method of equigravitating elements was developed for studying gravitational fields and calculating potential energy for the bodies of more complicated shapes \citep{Kond1989},~\citep{Kond2001},~\citep{Kond2003},~\citep{Kond2007}, with help of this method new results in the theory of potential were obtained.

Calculating \textit{mutual gravitational energy} of the bodies $W_{mut}$ represents a separate and important class of problems in the theory of potential. In the current paper we consider a problem of mutual potential energy for the system of two homogeneous prolate spheroids, whose symmetry axes lie on the same line. Such binary system of spheroids is a good approximation for several astrophysical problems. Examples include systems of celestial bodies with double spin-orbital resonance (in particular, a double planet Pluton-Charon or some binary asteroids with synchronized rotation), also systems of close binary stars and some variants of the Roche problem \citep{Chandra1973}. However, yet until recently the expressions for the mutual gravitational energy for two prolate spheroids has not been derived. In this paper we solve this problem and find the mutual gravitational energy for prolate spheroids $W_{mut}$ in a finite analytical form through elementary functions.

\section{MUTUAL GRAVITATIONAL ENERGY OF THE SPHEROIDS PROLATED ALONG THE SAME LINE}\label{lab:muten}In cylindrical coordinates  $\left( r,x_{3}\right)$ an equation for prolate spheroid's surface has the form
\begin{equation}\label{6}
 \displaystyle \frac{r^{2}}{a_{1}^{2}} + \frac{x_{3}^{2}}{c^{2}} =1, \quad c\geq a_{1}.
\end{equation}
Let us consider a system of two gravitating homogeneous prolate spheroids with the semiaxes $\left( c_{1} \geq a_{1}\right)$ and $\left( c_{2} \geq a_{2}\right)$. These spheroids are situated in such a way, that their axes of symmetry coincide with the axis $Ox_3$ (fig.~\hyperref[fig:fig1]{1}), and the distance between centers of the figures is $O_{1}O_{2}=h$. For given densities $\rho_{1}$ and $\rho_{2}$ masses of the bodies will be $M_{1}=\displaystyle \frac{4}{3} \pi a_{1}^{2} c_{1} \rho_{1} $ and $M_{2}=\displaystyle \frac{4}{3} \pi a_{2}^{2} c_{2} \rho_{2}$.

\begin{adjustwidth}{-1.5in}{0in}

\begin{center}\label{fig:fig1}
\includegraphics[width=0.5\textwidth]{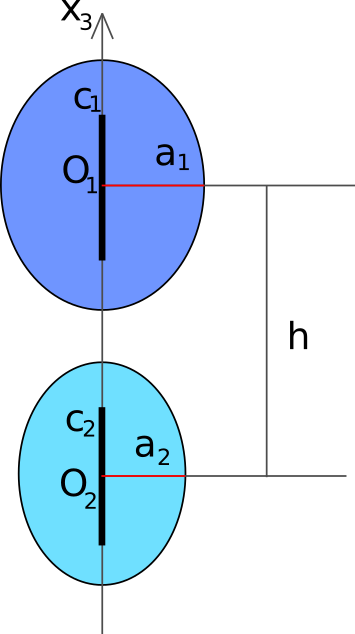}\end{center}\noindent\color{Gray} \textbf {Figure 1.}. Cross-sections of two aligned prolate spheroids. Their semiaxes and the distance $h$ between their centers $O_1O_2$ are showed. Focal equigravitating rods with lengths $L_{1}=2R_{1}$ and $L_{2}=2R_{2}$ are highlighted inside each of the spheroids.
\end{adjustwidth}
\vspace{.7cm}

As it is shown on the fig.~\hyperref[fig:fig1]{1}, the center $O_1$ of the first spheroid is chosen as the origin of the reference frame. In this case \citep[p. 255]{Kond2007} density of the focal equigravitating rod (with the length $L_1=2R_1$) for the upper prolate spheroid will be equal to 
\begin{equation}\label{7}
 \mu_{1} \left( x_{3} \right) =  \displaystyle  \frac{3 M_{1} }{4 R_{1}} \left( 1 - \frac{x_{3}^{2}}{R^{1}}\right), \quad R_{1}=\sqrt{c_{1}^{2} - a_{1}^{2}}, \quad -R_{1}\leq x_{3} \leq R_{1},
\end{equation}
and an external potential for the lower prolate spheroid on the symmetry axis $Ox_3$ will be \citep[p. 163]{Kond2007}
\begin{equation}\label{8}
 \varphi_{2} \left( x_{3} \right) = \displaystyle \frac{3 G M_{2} }{4 R_{2}} \left\lbrace \frac{2 \left( x_{3} + h \right)}{R_{2}} + \left(1 - \frac{\left( x_{3} + h \right)^{2}}{R_{2}^{2}}\right) \ln\frac{x_{3} + h + R_{2}}{x_{3} + h - R_{2}} \right\rbrace .
\end{equation}
Here we use designation
\begin{equation}\label{9}
 R_{2}=\sqrt{c_{2}^{2} - a_{2}^{2}}.
\end{equation}

It must be emphasized, that the method of equigravitating elements allows us to replace an external potential for any three-dimensional homogeneous prolate spheroid by a potential of its one-dimensional focal equigravitating rod with a parabolic density law (\ref{7}), and this replacement eases further calculations sufficiently. Integrating the potential (\ref{8}) over the rod, corresponding to the upper spheroid (look at the formulae (\ref{4})), will give us mutual gravitational energy for the system of two spheroids \citep{Kond2007}
\begin{equation}\label{10}
W_{mut} = - \int \limits_{-R_{1}}^{R_{1}}\mu_{1} \left(x_{3}\right) \varphi_{2} \left(x_{3}\right) d x_{3}.
\end{equation}

The integral (\ref{10}) can be represented in a finite analytical form. To accomplish this, let us change the variable $z=\displaystyle \frac{x_{3}}{R_{1}}$ and use designations
\begin{equation}\label{11}
x=\displaystyle \frac{h}{R_{1}};\quad n=\frac{R_{1}}{R_{2}}=\frac{c_{1} e_{1}}{c_{2} e_{2}};\quad \alpha=x+\frac{1}{n};\quad \beta=x-\frac{1}{n},
\end{equation}where $e_1 =\displaystyle \sqrt{1-\frac{a_{1}^{2}}{c_{1}^{2}}}$ and $e_2 = \displaystyle \sqrt{1-\frac{a_{2}^{2}}{c_{2}^{2}}}$ are the eccentricities of the spheroids. Then the integral (\ref{10}) takes the form
\begin{equation}\label{12}
W_{mut} = \displaystyle -\frac{9}{16} \frac{G M_{1} M_{2}}{h} x n \int \limits_{-1}^{1} \left( 1-z^{2} \right) \left\lbrace 2n\left(z+x\right)+ \left[ 1-n^{2} \left(z+x\right)^{2} \right] \ln\frac{z+\alpha}{z+\beta} \right\rbrace dz.
\end{equation}
Calculating this integral, after many transformations we get an expression for the mutual energy
\begin{equation}\label{13}
 W_{mut} = \displaystyle -\frac{G M_{1} M_{2}}{h} \Phi \left( n,x\right) .
\end{equation}Here we introduce designations:
\begin{equation}\label{14}
 \displaystyle \Phi \left( n,x\right) = f_{1} \ln\frac{n\left( x+1\right)+1}{n\left( x+1\right)-1} + 
 f_{2} \ln\frac{n\left( x-1\right)+1}{n\left( x-1\right)-1} +
 f_{3} \ln\frac{n\left( x-1\right)-1}{n\left( x+1\right)-1} +f_{4}
\end{equation}
\begin{equation}\label{15}
\begin{array}{lcl}
f_{1} =  \displaystyle -\frac{3x}{160n^{2}} \left( n \left(x + 1\right) + 1 \right)^{3} \left( 4 - 12 n + 2 n x + 4 n^{2} + 3 n x + 3 n^{2} x - n^{2} x^{2}\right) \\ \\
f_{2} =  \displaystyle \frac{3x}{160n^{2}} \left( n \left(x - 1\right) + 1 \right)^{3} \left( 4 + 12 n + 2 n x + 4 n^{2} + 3 n x - 3 n^{2} x - n^{2} x^{2} \right) \\ \\
f_{3} =  \displaystyle \frac{3x}{20n^{2}} \left( 1-5n^{2}+5n^{2} x^{2} \right) \\ \\
f_{4} =  \displaystyle \frac{3x^{2}}{40} \left( 11 + 11 n^{2} + n^{2} x^{2} \right)
\end{array}
\end{equation}The dependence of an auxiliary function $\Phi\left( n,x\right)$ (\ref{14}) on the distance between the centers of the bodies is shown in the figure~\hyperref[fig:fig2]{2}. We can see, that the mutual potential energy for two spheroids rapidly decreases in case of close distances, while for larger distances the decreasing slows down.
\begin{adjustwidth}{-1.5in}{0in}

\begin{center}\label{fig:fig2}
\includegraphics[width=0.5\textwidth]{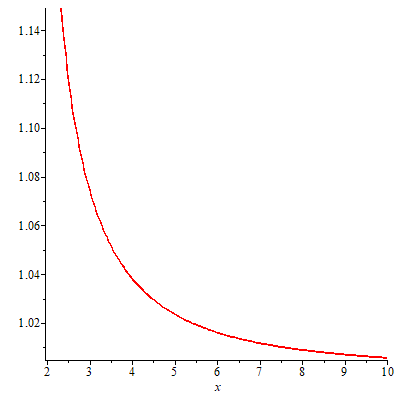}\end{center}\noindent\color{Gray} \textbf {Figure 2.} The dependence of the normalized mutual energy $\Phi \left( n,x\right)$ for two prolate spheroids on the distance between their centers $x=\displaystyle \frac{h}{R_{1}}$, according to the formulae (\ref{14}) and (\ref{15}) for $n=1$.
\end{adjustwidth}
\vspace{.7cm}

It can be noticed, that the expression (\ref{14}) may be rewritten as power series in eccentricities of the spheroids $e_1$ and $e_2$. For this purpose, let us introduce volume-equivalent radii $\tilde{R_1}$ and $\tilde{R_2}$ and suppose, that in a case of change in eccentricity the spheroids retain their masses and volumes, i.e. $\tilde{R}_{1}^{3} = a_{1}^{2} c_{1} $, $\tilde{R}_{2}^{3} = a_{2}^{2} c_{2}$. Then, obviously,
\begin{equation}\label{16}
 R_{1}=\displaystyle \frac{e_{1} \tilde{R}_{1}}{\left( 1-e_{1}^{2}\right)^{\frac{1}{3}}};\quad R_{2}=\displaystyle \frac{e_{2} \tilde{R}_{2}}{\left( 1-e_{2}^{2}\right)^{\frac{1}{3}}}. 
\end{equation}Introducing normalized quantities $x_{s} =\displaystyle  \frac{h}{\tilde{R}_{1}}$, $n_{s} = \displaystyle \frac{\tilde{R}_{1}}{\tilde{R}_{2}}$ we obtain
\begin{equation}\label{17}
x=x_{s} \displaystyle \frac{\left( 1-e_{1}^{2}\right)^{\frac{1}{3}}}{e_{1}};\quad n=n_{s} \frac{e_{1} \left( 1-e_{2}^{2}\right)^{\frac{1}{3}}}{e_{2} \left( 1-e_{1}^{2}\right)^{\frac{1}{3}}}. 
\end{equation}After calculations with accuracy up to the terms of small order $e_{1}^{5}$ and $e_{2}^{5}$ included, we find a new expression for the mutual gravitational energy:
\begin{equation}\label{18}
 W_{mut} = \displaystyle -\frac{G M_{1} M_{2}}{h} \left\lbrace 1 + \frac{e_{1}^{2}}{5 x_{s}} +\frac{e_{2}^{2}}{5 x_{s} n_{s}^{2}} + \frac{e_{1}^{2} e_{2}^{2}}{5 x_{s}^{2} n_{s}^{2}} \left( \frac{2}{3}+\frac{1}{5x_{s}}+\frac{1}{x_{s}^{2}}\right) + ... \right\rbrace. 
\end{equation}

Function $\Phi \left( n,x\right)$ represents the mutual energy $W=\displaystyle \frac{-W_{mut} h}{G M_{1} M_{2}}$ of two homogeneous prolate spheroids, normalized to the standard value of energy for two spheres (or point masses) of the same mass. Figs. ~\hyperref[fig:fig2]{2} and ~\hyperref[fig:fig3]{3} shows, that the function $\Phi \left( n,x\right)$ rapidly decreases with the increasing distance between the centers of the spheroids but is limited from below, always being larger than unity.
\begin{equation}\label{19}
 \Phi \left( n,x\right) > 1. 
\end{equation}
\begin{adjustwidth}{-1.5in}{0in}

\begin{center}\label{fig:fig3}
\includegraphics[width=0.5\textwidth]{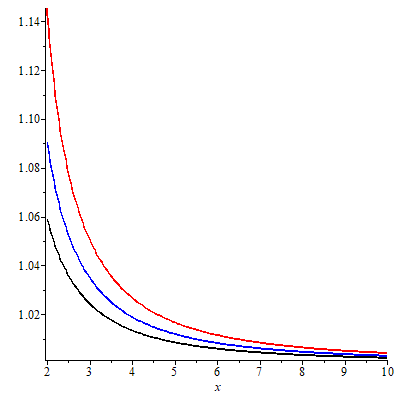}\end{center}\noindent\color{Gray} \textbf {Figure 3.} The dependence of the function $\Phi \left( n,x\right)$ for different $n$ on the normalized distance $x=\displaystyle \frac{h}{R_{1}}$. Curves are shown for $n=1$ (red), $n=1.5$ (blue), $n=5$ (black).
\end{adjustwidth}
\vspace{.7cm}

It means, that the mutual energy for two homogeneous prolate spheroids with their symmetry axes lying on the same line will be always larger than the mutual energy for two equivalent spheres or point masses.

Differentiating the mutual potential (\ref{13}) we can find a gravitational force between two prolate spheroids:
\begin{equation}\label{20}
 F \left( n,x\right) =\displaystyle  \frac{G M_{1} M_{2}}{h^{2}} \left( \Phi \left( n,x\right) - x \frac{d\Phi \left( n,x\right)}{dx} \right) .
\end{equation}
Substituting $\Phi \left( n,x\right)$ from (\ref{14}), after many transformations we obtain
\begin{equation}\label{21}
\begin{array}{lcl}
F \left( n,x\right) = \displaystyle \frac{G M_{1} M_{2}}{h^{2}} \cdot \\ \\
 \displaystyle \left\lbrace \left[ -\frac{3x^{2}}{32n} \left( n x + n + 1 \right)^{2} \left( n^{2} x^{2} - 2n^{2} x - 3n^{2} -2n x + 6n - 3 \right) \right] \ln\frac{n\left( x+1\right)+1}{n\left( x+1\right)-1} + \right. \\ \\ 
\displaystyle +\left[ \frac{3x^{2}}{32n} \left( n x - n + 1 \right)^{2} \left( n^{2} x^{2} + 2n^{2} x - 3n^{2} -2n x - 6n - 3 \right) \right] \ln\frac{n\left( x-1\right)+1}{n\left( x-1\right)-1} - \\ \\
\displaystyle \left.-\frac{3x^{3}}{2} \ln\frac{n\left( x-1\right)-1}{n\left( x+1\right)-1} -\frac{3x^{2}}{8} \left( n^{2} x^{2} + 3n^{2} + 3 \right) \right\rbrace .
\end{array}
\end{equation}
For the example figure ~\hyperref[fig:fig4]{4} shows the dependence of the attraction force on the distance between centers of two identical prolate spheroids. We can see, that the attraction force between prolate spheroids is always larger than the force between two spheres (or point masses) of the same masses.
\begin{adjustwidth}{-1.5in}{0in}

\begin{center}\label{fig:fig4}
\includegraphics[width=0.5\textwidth]{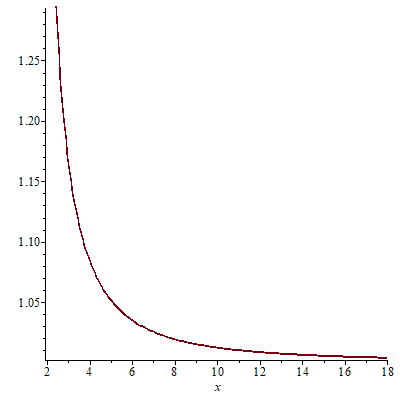}\end{center}\noindent\color{Gray} \textbf {Figure 4.} The dependence of modulus of the normalized gravitational force $F \left( n,x\right) =\displaystyle \frac{G M_{1} M_{2}}{h^{2}}$ on the distance $x=\displaystyle \frac{h}{R_{1}}$ between the centers of two homogeneous aligned prolate spheroids. Calculations are performed using the formula (\ref{21}) for $n=1$.
\end{adjustwidth}
\vspace{.7cm}

\section{DISCUSSION AND CONCLUSIONS}\label{lab:resume}
In the current paper we set and solved the problem of mutual gravitational energy $W_{mut}$ for two homogeneous prolate spheroids, orientated so that their symmetry axes lie on the same line. To solve the problem, we applied the method of equigravitating elements developed by Kondratyev. In this method external potentials of three-dimensional homogeneous spheroids are represented by potentials of one-dimensional inhomogeneous focal rods, thus the problem solving is reduced to integrating of the first rod's potential over the length of the second one. As a result, it becomes possible to obtain the mutual energy $W_{mut}$ for two prolate spheroids and the attraction force between them in a finite analytical form expressed through elementary functions. More than that, the function $W_{mut}$ can be represented by the power series in eccentricities of the spheroids.

The expression for the mutual energy obtained here may be used for many astrophysical tasks, for example, for studying dynamics of close binary stars and asteroids, contact pairs in particular, as also double planets like the system Pluton-Charon.

It must be noticed, that among the variety of binary asteroids there was no any known pair with doubly synchronized rotation. As the tidal forces are inversely proportional to the cube of the distances, tidal captures should be sought among pairs of asteroids that are very close to each other. 

However, a pair of asteroids with synchronized rotation was discovered recently \citep{Taylor2019}. It has an assigned number \textbf{(190166) 2005 UP156}. This binary system is especially amazing because it consists of asteroids that are in a state of double tidal capture. Observations showed, that the asteroids of this system have approximately same masses and congruent prolate shapes. The study of this binary asteroid's dynamics is a matter of a separate paper.

\end{document}